\documentclass[letterpaper,preprint,12pt,tightenlines,amsmath,amssymb,floatfix,nofootinbib,aps,raggedbottom]{revtex4-1}

\usepackage{graphicx}
\usepackage{bm}
\usepackage{url,multirow,xcolor}
\usepackage[font=footnotesize,margin=24pt,justification=centerlast]{caption}

\begin{document}

\title{ Estimation of nuclear matrix elements of double-$\bm{\beta}$ decay from shell model and quasiparticle random-phase approximation \\ \vspace{5pt} }

\author{J.\ Terasaki\footnote{jun.terasaki@cvut.cz} \\ \vspace{0pt}}
\affiliation{ Institute of Experimental and Applied Physics\hbox{,} Czech Technical University in Prague, Husova 240/5, 110\hspace{3pt}00 Prague 1, Czech Republic}
\author{Y.\ Iwata \\ \vspace{0pt}}
\affiliation{Faculty of Chemistry\hbox{,} Materials and Bioengineering\hbox{,} Kansai University, Yamatemachi 3-3-35\hbox{,} Suita\hbox{,} Osaka 564-8680, Japan}

\author{}
\affiliation{}

\begin{abstract}
The nuclear matrix element (NME) of the neutrinoless double-$\beta$ ($0\nu\beta\beta$) decay is an essential input for determining the neutrino effective mass, if the half-life of this decay is measured. The reliable calculation of this NME has been a long-standing problem because of the diversity of the predicted values of the NME depending on the calculation method. In this paper, we focus on the shell model and the QRPA. The shell model have a rich amount of the many-particle many-hole correlations, and the QRPA can obtain the convergence of the result of calculation with respect to the extension of the single-particle space. 
It is difficult for the shell model to obtain the convergence of the $0\nu\beta\beta$ NME with respect to the valence single-particle space. The many-body correlations of the QRPA are insufficient depending on nuclei. We propose a new method to modify phenomenologically the results of the shell model and the QRPA compensating the insufficient point of each method by using the information of other method complementarily. Extrapolations of the components of the $0\nu\beta\beta$ NME of the shell model are made toward a very large valence single-particle space. 
We introduce a modification factor to the components of the $0\nu\beta\beta$ NME of the QRPA. 
Our modification method  gives similar values of the $0\nu\beta\beta$ NME of the two methods for $^{48}$Ca. The NME of the two-neutrino double-$\beta$ decay is also modified in a similar but simpler manner, and the consistency of the two methods is improved.
\end{abstract}
\maketitle
\section{Introduction}
\label{sec:introduction}
The neutrinoless double-$\beta$ ($0\nu\beta\beta$) decay has been studied intensively by many researchers since the prediction by Ref.~\cite{Fur39} as one of the clues for new physics. If this decay is found, it implies that the neutrino is a Majorana particle, and the lepton number is not conserved. In addition, if the half-life is measured, it is possible to determine the effective neutrino mass, which is also called Majorana mass. The neutrino had been thought to be a particle either massless or very light for many decades until the discovery of the neutrino oscillation \cite{Fuk98,Ahm02,Egu03,Ali05}, which has proven the massiveness. Currently the determination of the mass scale of the neutrino is one of the major subjects of the neutrino physics. 

The $0\nu\beta\beta$ decay offers one of the limited methods to determine this mass scale. Large-scale experimental projects around twenty \cite{CANDLES,GERDA,Majorana_Demonstrator,LEGEND,SuperNEMO,ZICOS,AMoRE,CROSS,CUORE,SNO+,Gan16,nEXO,NEXT,CUPID,LUCIFER,COBRA,DARWIN,LUX-ZEPLIN} are in progress for observing the $0\nu\beta\beta$ decay with the assumption of the Majorana nature of the neutrino. For other methods to use nuclei, see Refs.~\cite{KATRIN,Project8,ECHo}. In the current status of no successful report of observing the $0\nu\beta\beta$ decay, the upper limit of the effective neutrino mass is deduced, e.g.~Ref.~\cite{Gan16}, from the performance of the detector system, the theoretically calculated nuclear matrix element (NME) and the phase-space factor originating from the emitted electrons. The NME is more difficult to calculate accurately than the phase-space factor because the nuclear wave functions are necessary. The lightest nucleus used for the experiments is $^{48}$Ca \cite{CANDLES}, and some approximation is essentially necessary for the wave functions of the involved nuclei; this necessity is obvious for the heavier candidate nuclei. 

A problem is well known that the calculated $0\nu\beta\beta$ NMEs are distributed in the range of a factor of 2$-$3 depending on the theoretical method to calculate the nuclear wave functions \cite{Fae12J,Eng17}. In particular, the shell model and the quasiparticle random-phase approximation (QRPA)\footnote{Most of studies with this approach use the proton-neutron QRPA, e.g.~\cite{Suh07}, which is called the QRPA for simplicity in this paper.}, which have been used intensively and historically, have a difference of the factor of two for several instances of the $0\nu\beta\beta$ decay \cite{Fae12J,Eng17}. The shell model is the diagonalization method of the many-body Hamiltonian, and many-particle many-hole (mpmh) correlations are included in the wave functions. We define the mpmh to be at least a two-particle two-hole (2p2h) configuration mixing. The QRPA is an approximation to obtain the transitions from the ground state to excited states, and this transition is limited to the two-quasiparticle creation and annihilation. The shell-model calculations are performed in many cases of these days with one major valence shell defined by the harmonic oscillator for the single particles, while the QRPA can use much larger single-particle space with a more realistic single-particle basis. The limit of the valence single-particle space or many-body correlations is mainly due to the technical limit of computation. The correct $0\nu\beta\beta$ NME should be confirmed by the convergence of the result with respect to the extension of the valence single-particle space and the mpmh components of the nuclear wave function. It is difficult to achieve this double convergence in spite of the quite remarkable development of the modern computers. The physical origin of this difficulty is the neutrino potential included in the $0\nu\beta\beta$ NME. This two-body potential has a singularity at the origin, therefore, a very large wave-function space is necessary. For the shell-model and similar calculations of the $0\nu\beta\beta$ NME for $^{48}$Ca, the two-major valence shell is the largest valence single-particle space ever used \cite{Iwa16,Jia17}. For the QRPA an extension called the renormalized QRPA \cite{Toi95,Sim97} has been investigated.

Another problem is the effective axial-vector current coupling denoted by $g_A$ for nuclei. The $0\nu\beta\beta$ NME is a linear combination of the Gamow-Teller (GT), Fermi, and tensor components, and the coefficient includes $g_A$. The tensor component is omitted in this paper because its contribution is small, e.g., \cite{Iwa16,Sim13}. The $g_A$ is equal to one in the quark-lepton level, e.g.~\cite{Com83}, while it is 1.27641(45)$_\mathrm{stat}$(33)$_\mathrm{sys}$ \cite{Mar19} for the neutron. This difference indicates that $g_A$ is affected by the many-body effects of the quarks. Thus, the corrections of the transition operator due to the many-body effects of the nucleons may be necessary, if the exact nuclear wave functions are available. The approximation of the wave functions causes another necessity of the effective $g_A$. 
There is a long history of theoretically deriving the effective $g_A$, e.g.~\cite{Tow87}.  The method to determine the effective $g_A$ is not yet established for the $0\nu\beta\beta$ decay. For recent attempts of theoretically deriving the effective transition operators of this decay, see Refs.~\cite{Cor20,Nov20} and references therein. 

In this paper, we propose a new approach to estimate the components of the $0\nu\beta\beta$ NME by modifying the results of the calculations of the shell model and the QRPA with the  compensation of the insufficient points of the two methods. It is difficult for the shell model to obtain the convergence of the $0\nu\beta\beta$ NME with respect to the extension of the valence single-particle space. The many-body correlations of the QRPA are insufficient depending on nuclei. 
Extrapolations of the components of the $0\nu\beta\beta$ NME of the shell model are made with respect to the energy representing the size of the valence single-particle space referring to the intermediate-state energy dependence of the components of the $0\nu\beta\beta$ NME of the QRPA. On the other hand, we introduce a modification factor to the components of the $0\nu\beta\beta$ NME of the QRPA from the comparisons of the charge-change strength functions of the experiments, the shell model, and the QRPA. This modification factor represents the mpmh effects missing in the QRPA.  $^{48}$Ca is used in this study because the shell-model calculations have been performed with the one- and two- major valence single-particle spaces \cite{Iwa16}. 
We discuss the GT and Fermi components separately for avoiding the involvement of the effective-$g_A$ problem.  

This paper is organized as follows; In Sec.~\ref{sec:basic_equations}, the basic equations used in this paper are summarized.  In Sec.~\ref{sec:GT0vbbNME_SM}, we discuss the modification of the GT component of the $0\nu\beta\beta$ NME of the shell model. Subsequently the modification of that of the QRPA is discussed in Sec.~\ref{sec:GT0vbbNME_QRPA}. Section \ref{sec:F0vbbNME} treats the Fermi component of the $0\nu\beta\beta$ NME. The $0\nu\beta\beta$ NME is calculated from the two components in Sec.~\ref{sec:0vbbNME}. In Sec.~\ref{sec:2vbbNME}, we discuss the modification of the NME of the two-neutrino double-$\beta$  ($2\nu\beta\beta$) decay. Section \ref{sec:summary} is the summary. 

\section{Basic equations}
\label{sec:basic_equations} 
Prior to the discussion, we summarize the basic equations and the definitions of the quantities relevant to the $0\nu\beta\beta$ decay. The probability of this decay, e.g.~\cite{Hax84,Doi85}, can be written 
\begin{eqnarray}
P_{0\nu} = \left| M^{(0\nu)}\right|^2 G_{0\nu}\left( \frac{\langle m_\nu \rangle}{m_e} \right)^2, 
\end{eqnarray}
where $M^{(0\nu)}$ is the NME of the $0\nu\beta\beta$ decay, and $G_{0\nu}$ denotes the phase-space factor, e.g.~\cite{Kot12}. The effective neutrino mass is denotd by $\langle m_\nu \rangle$, and $m_e$ is the electron mass. 
The half-life is inversely proportional to $P_{0\nu}$. 
The $M^{(0\nu)}$ discussed in this paper is calculated by 
\begin{eqnarray}
M^{(0\nu)} = M^{(0\nu)}_\mathrm{GT} - \left(\frac{g_V}{g_A}\right)^2 M^{(0\nu)}_\mathrm{F}. \label{eq:m0vbb}
\end{eqnarray}
$M^{(0\nu)}_\mathrm{GT}$ and $M^{(0\nu)}_\mathrm{F}$ denote the GT and the Fermi components, respectively. The constant $g_V$ is the vector-current coupling.  
The GT component with the closure approximation, e.g.~\cite{Hor10,Sim11}, is given by 
\begin{eqnarray}
M^{(0\nu)}_\mathrm{GT} &=& 
\sum_{pnp^\prime n^\prime}
\langle pp^\prime| V^{(0\nu)}_\mathrm{GT}(r;{\bar{E}}_B) |nn^\prime\rangle
\langle F| c^\dagger _p c_n c^\dagger_{p^\prime} c_{n^\prime} | I \rangle,
\label{eq:m0vbbGT}
\end{eqnarray}
where $p$ and $p^\prime$ ($n$ and $n^\prime$) denote the proton (neutron) states, and $c_i^\dagger$ is the creation operator of the single-particle state $i$ ($p$ or $n$); the annihilation operator is $c_i$.  $V^{(0\nu)}_\mathrm{GT}(r;{\bar{E}}_B)$ is the double GT transition operator of the $0\nu\beta\beta$ decay with the two-nucleon distance $r$ and 
${\bar{E}}_B$ to be the average energy of the intermediate states $|B\rangle$, which are the virtual states between the first and second $\beta$ decays.  
$|I\rangle$ and $|F\rangle$ denote the initial and final states of the decay, respectively, and the ground states are used. 
$V^{(0\nu)}_\mathrm{GT}(r;{\bar{E}}_B)$ is defined by
\begin{eqnarray}
V^{(0\nu)}_\mathrm{GT}(r;\bar{E}_B) = h_+(r,\bar{E}_B) \bm{\sigma}(1)\cdot\bm{\sigma}(2) \tau^-(1) \tau^-(2), \label{eq:vpotGT}
\end{eqnarray}
where $\bm{\sigma}$ is the spin Pauli operator, and $\tau^-$ indicates the operator changing the neutron to the proton. The arguments 1 and 2 distinguish the two particles operated. The function $h_+(r,\bar{E}_B)$ is the neutrino potential, of which behavior is similar to that of the Coulomb potential with the singularity at $r$ = 0. For the equation of $h_+(r,\bar{E}_B)$, see Ref.~\cite{Eng17,Ter15}. Equation (\ref{eq:m0vbbGT}) is  used by the shell model. 

The QRPA uses two sets of the intermediate states due to the feature of the approximation. One is $|B_I\rangle$ obtained on the basis of $|I\rangle$, and another is $|B_F\rangle$ based on $|F\rangle$.   
The equation of $M^{(0\nu)}$ for the QRPA approach reads 
\begin{eqnarray}
M^{(0\nu)}_\mathrm{GT} (\mathrm{QRPA}) &=& 
\sum_{B_{F}\, B_{I}} \sum_{pnp^\prime n^\prime}
\langle pp^\prime| V^{(0\nu)}_\mathrm{GT}(r;{\bar{E}}_B) |nn^\prime\rangle
\langle F| c^\dagger _p c_n | B_F \rangle 
\langle B_F | B_I \rangle \langle B_I | c^\dagger_{p^\prime} c_{n^\prime} | I \rangle. 
\label{eq:m0vbbGTQRPA}
\end{eqnarray}
Below, we refer to this one as $M^{(0\nu)}_\mathrm{GT}$ of the QRPA. The intermediate states are explicitly used because the QRPA is suitable to the calculation of the transition-density matrix elements 
$\langle F| c^\dagger _p c_n | B_F \rangle$ and 
$\langle B_I | c^\dagger_{p^\prime} c_{n^\prime} | I \rangle$. 
For the calculation of the overlap $\langle B_F|B_I\rangle$, see Refs.~\cite{Ter12,Ter13}. 
The equations for the Fermi component are the same as Eqs.~(\ref{eq:m0vbbGT})$-$(\ref{eq:m0vbbGTQRPA}) except that the double-spin operator $\bm{\sigma}(1)\cdot\bm{\sigma}(2)$ is not used. 

\section{Modification of GT component of $\bm{0\nu\beta\beta}$ NME of shell model}
\label{sec:GT0vbbNME_SM}

\subsection{Method with the help of experimental strength function}
\label{sec:exp_str_fn}
Our study is based on the shell-model result of Ref.~\cite{Iwa16} and the QRPA result of Ref.~\cite{Ter18} for two reasons. One is that Ref.~\cite{Iwa16} includes one- and two- major valence shell calculations. The extension of the valence single-particle space is important to our study. Another reason is that the energy dependences of the charge-change strength functions of the two methods are similar in the energy region up to the GT giant resonance; see Refs.~\cite{Iwa16,Ter18}. Thus, the physical effects of the interactions used in the two calculations are similar. Probably this similarity is expected because both interactions are  phenomenological. The shell-model results used in this paper were obtained using the interactions GXPF1B and SDPFMU-DB \cite{Iwa16}. For other calculations of $^{48}$Ca using the shell model or methods related to this model, see Refs.~\cite{Jia17,Nov20,Hax84,Hor10,Men14,Cor20PRC,Kos20,Yao20,Bel21} and the references cited therein. See also Refs.~\cite{Sim13,Suh93} for other QRPA calculations of this nucleus. 

Here, we describe the technical aspects of our QRPA calculation. 
The particle-hole interaction is the Skyrme (parameter set SkM$^\ast$ \cite{Bar82}), and the pairing interaction is the contact interaction with no density dependence.  The strength of the pairing interaction was determined so as to reproduce the pairing gap deduced from the mass data by the three-point formula \cite{Boh69} with a very low cutoff occupation probability for the canonical single-particle states in the paired case or a very high cutoff energy in the unpaired case in the Hartree-Fock-Bogoliubov calculations. Namely, those are the pairing interactions for the like-particles. The isoscalar and isovector proton-neutron pairing interactions were also used for the QRPA calculation; for these interactions, see Ref.~\cite{Ter20} and references therein. The calculations are performed in the M scheme. The number of the single-particle levels used for the QRPA calculation is around 1700 for each of the protons and neutrons; that is the valence single-particle space. The total number of QRPA solutions is nearly 600000 for each of $^{48}$Ca and $^{48}$Ti. For detail of the calculation, see Ref.~\cite{Ter18}. 

\begin{figure}[t]
\centering
\includegraphics[width=0.5\columnwidth]{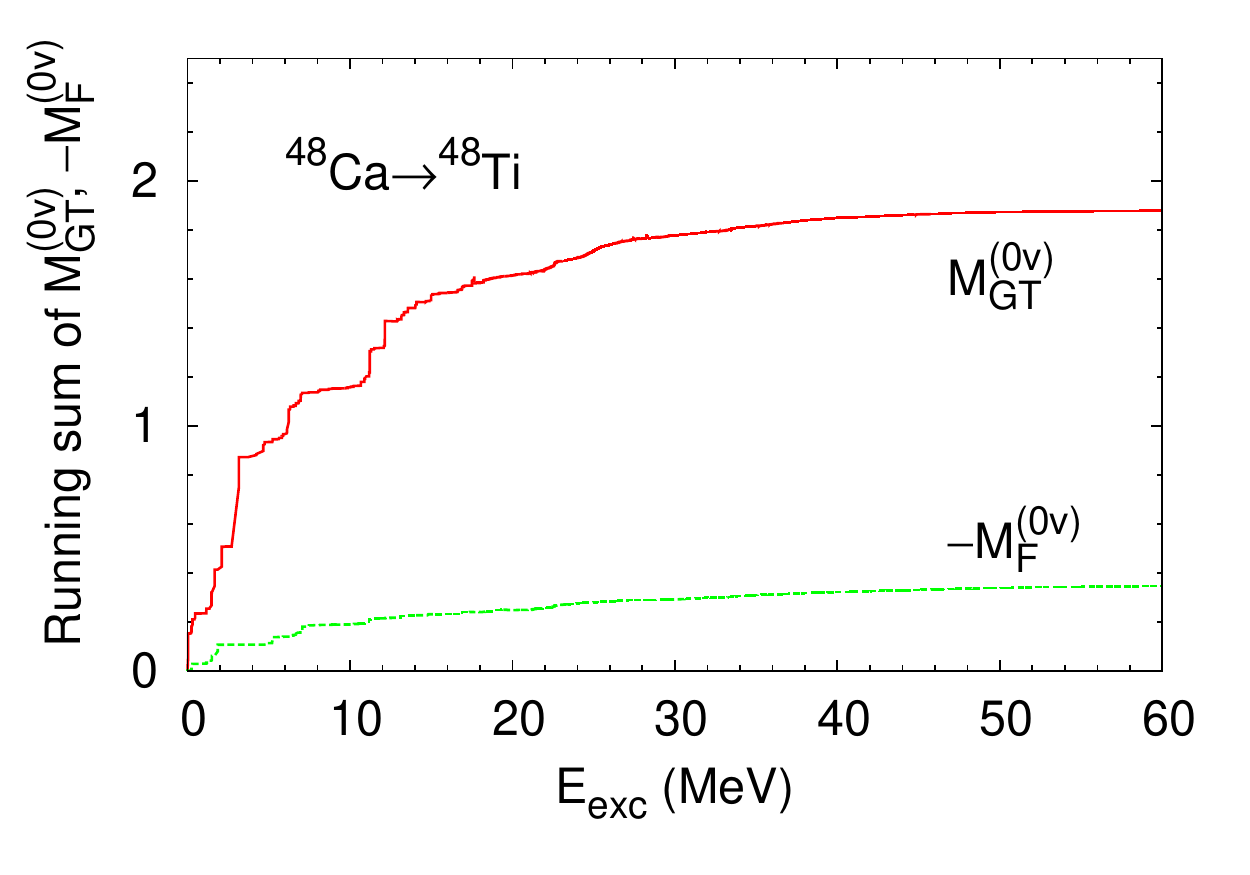}
\vspace{-10pt}
\caption{ \protect \label{fig:rs0vbbGTF} \baselineskip=13pt Running sums of $M^{(0\nu)}_\mathrm{GT}$ and $M^{(0\nu)}_\mathrm{F}$ of $^{48}$Ca$\rightarrow$$^{48}$Ti obtained by the QRPA \cite{Ter18} as functions of the excitation energy $E_\mathrm{exc}$ of the intermediate nucleus $^{48}$Sc. The actual calculation has been performed up to around 75 MeV. No quenching factor is used. }
\end{figure}

We obtained the running sums (cumulation) of $M^{(0\nu)}_\mathrm{GT}$ and $M^{(0\nu)}_\mathrm{F}$ from the QRPA solutions for $^{48}$Ca$\rightarrow$$^{48}$Ti \cite{Ter18} as shown in Fig.~\ref{fig:rs0vbbGTF}. 
The horizontal axis indicates the excitation energy $E_\mathrm{exc}$ of the intermediate nucleus $^{48}$Sc. The two NME components converge around $E_\mathrm{exc}$ = 50 MeV. The first step of our approach is to compare the energy dependence of this running sum with the results of the shell-model calculation with the truncated valence single-particle spaces. 
The shell model is usually applied to the $0\nu\beta\beta$-NME calculation without the intermediate states. It is necessary for our discussion to consider what maximum $E_\mathrm{exc}$ the shell-model calculations have effectively. Suppose that the maximum one-particle one-hole (1p1h) energy is 5 MeV, and the two involved single-particle levels have a ten-fold degeneracy; then, the 10p10h energy is 50 MeV under the assumption that the lower level is fully occupied in the lowest-energy configuration. The comparison of the NME of this shell model and the QRPA running NME at $E_\mathrm{exc}$ = 50 MeV would not be useful because the low-order particle-hole excitations of 50 MeV are not included in the shell-model calculation. Thus, it is not a trivial question what energy of the shell model corresponds to $E_\mathrm{exc}$ of the QRPA. A method to answer this question is to refer to the charge-change transition strength with the angular momentum and the parity of $J^{\pi}$ = $1^+$ of the shell model and the experimental data. 

\begin{figure}[]
\includegraphics[width=0.95\columnwidth]{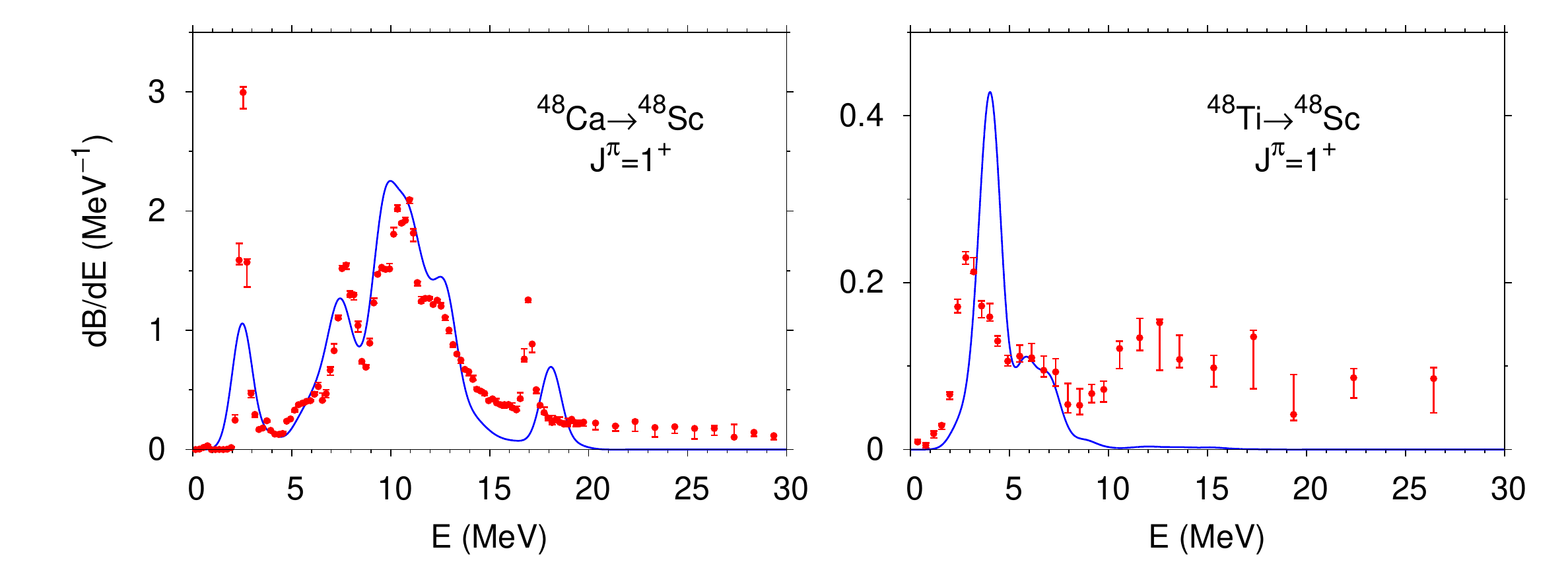}
\vspace{-5pt}
\caption{ \protect \label{fig:strfn48Ca48TiIwata} \baselineskip=13pt Charge-change strength function $dB/dE$ with $J^\pi=1^+$ based on a shell-model calculation (blue curves) accounting
for the \textit{pf} valence shell (interaction GXPF1B) and experimental data (red bars) \cite{Yak09} as functions of excitation energy of $^{48}$Sc. The left (right) figure shows the strength function from $^{48}$Ca ($^{48}$Ti) to $^{48}$Sc. For the calculation, GT$^-$ (GT$^+$) operator $\bm{\sigma}\tau^-$ ($\bm{\sigma}\tau^{- \dagger}$)  with a quenching factor 0.77 is used for the left (right) figure. For the experimental data, the possibility is pointed out  \cite{Yak09} that the isovector spin monopole operator is involved in addition to the GT operator.
The Gaussian function is applied to the calculation with its square root of the variance: $\sigma$ = 0.510 MeV. The experimental data have the full widths at half maximum of energy resolution 0.2 MeV corresponding to $\sigma$ = 0.085 MeV in the left figure and 0.4 MeV corresponding to $\sigma$ = 0.170 MeV in the right figure.  }
\end{figure}

For the shell model, the authors of Ref.~\cite{Iwa15} fitted the experimental charge-change strength function of $^{48}$Ca and $^{48}$Ti, as shown by Fig.~\ref{fig:strfn48Ca48TiIwata}, by a quenching factor of 0.77 to the GT operator in the \textit{pf} valence shell calculation; the quenching factor to the strength function is 0.59. 
In our observation, their fitting is good up to 13 MeV for $^{48}$Ca and 7.5 MeV for $^{48}$Ti. The NME of the double-$\beta$ decay needs the products of the transition densities of the $^{48}$Ca$\rightarrow$$^{48}$Sc and $^{48}$Ti$\rightarrow$$^{48}$Sc. Thus, it is inferred that the shell model with the \textit{pf} valence shell is reliable for the components of $M^{(0\nu)}_\mathrm{GT}$ of $^{48}$Ca up to $E_\mathrm{exc}$ = 7.5 MeV. Since the valence single-particle space is essential to the determination of the reliable energy region, we assume that this region is not changed appreciably for other not-very-high angular momenta and parity, as long as the same valence single-particle space is used. 

A question arises why the quenching factor is necessary, if the shell model is reliable. One of the reasons is the necessity to add the isovector spin monopole (IVSM) operator, i.e., the GT operator multiplied by $r_1^2$ ($r_1$ is the radial variable of a nucleon), to the transition operator as pointed out in Ref.~\cite{Yak09}. The discussion in this section is not affected by the IVSM operator. However, the following point is worthy of noting; if the IVSM operator is used, the missing transition strength in $E_\mathrm{exc}$ $>$ 7.5 MeV in $^{48}$Ti$\rightarrow$$^{48}$Sc would not be reproduced in the present shell-model calculations with the \textit{sdpf} valence shell. The two-major-shell jump is necessary for activating the high-energy components of the IVSM operator \cite{Ter18,Min16}, thus, the \textit{sdpf} valence shell is not sufficient. The analogous discussion is applied to the tail region of the strength function of $^{48}$Ca$\rightarrow$$^{48}$Sc. 
The possibility is also noted that the transition operator should be modified, not as a compensation of approximation, by the many-body effects as mentioned in the introduction in relation to the effective $g_A$. 

The $M^{(0\nu)}_\mathrm{GT}$ of the QRPA is 1.14 at $E_\mathrm{exc}$ = 7.5 MeV, and the converged value is 1.88; they are summarized in Table \ref{tab:0vbbGTQRPA}. The shell model should include the 1p1h correlations of the QRPA in any energy region, if the valence single-particle space is sufficiently large. Thus, the increasing behavior of the $M^{(0\nu)}_\mathrm{GT}$ and $M^{(0\nu)}_\mathrm{F}$ of the QRPA should be included in that very large shell-model calculation. We estimate the effect of the missing valence single-particle space in the current shell model by the ratio of the converged $M^{(0\nu)}_\mathrm{GT}$ of the QRPA to the running sum up to $E_\mathrm{exc}$ = 7.5 MeV. 
The components of $0\nu\beta\beta$ NME of the shell model are shown in Table \ref{tab:0vbbGTFSM}. We use the average values of the two short-range correlations (SRC): CD-Bonn and Argonne. The average $M^{(0\nu)}_\mathrm{GT}$ of the \textit{pf} valence-shell calculation is 0.77, and that of the $sdpf$ valence shell is 1.00. The converged $M^{(0\nu)}_\mathrm{GT}$ of the shell model is estimated to be the shell-model value multiplied by the increasing ratio of the QRPA, that is, 
\begin{eqnarray}
 0.77 \cdot (1.88/1.14) = 1.27. \label{eq:est0vbbGTSM_strfn}
\end{eqnarray}
The underlying assumption for this estimation is that the effects of the mpmh correlations increase with the same ratio of the 1p1h effects. This assumption is justified in another method discussed in the next section.

\begin{table}[t]
\caption{\label{tab:0vbbGTQRPA} \baselineskip=13pt Partial $M^{(0\nu)}_\mathrm{GT}$ of the QRPA at $E_\mathrm{exc}$ = 7.5 MeV in the running sum and the converged value for $^{48}$Ca. }
\centering
\begin{tabular}{ cc }
\hline \\[-11pt]
$\ E_\mathrm{exc}$ (MeV)\quad &  \quad$M^{(0\nu)}_\mathrm{GT}$(QRPA)
\quad \\[2pt]
\hline \\[-12pt]
     7.5         & 1.14  \\
   $\sim$ 70 & 1.88  \\
\hline
\end{tabular}
\vspace{6pt}
\end{table}

\begin{table}[]
\caption{\label{tab:0vbbGTFSM} \baselineskip=13pt Components of $M^{(0\nu)}_\mathrm{GT}$ and $M^{(0\nu)}_\mathrm{F}$ of the shell model for $^{48}$Ca \cite{Iwa16}. We show the results with the interactions GXPF1B for the \textit{pf} valence shell and SDPFMU-DB for the \textit{sdpf} valence shell; only the results necessary for our discussion are referred to. SRC denotes the short-range correlation; CD-Bonn and Argonne were used. No quenching factor is used. }
\centering
\begin{tabular}{ cccccc }
\hline \\[-12pt]
   \multirow{2}{*}{ \quad SRC \quad} &  \multicolumn{2}{c}{ \textit{pf} } & &  \multicolumn{2}{c}{ \textit{sdpf} }  \\[1pt]
  \cline{2-3} \cline{5-6} \\[-11pt]
 & \quad $M^{(0\nu)}_\mathrm{GT}$ \quad & \quad $M^{(0\nu)}_\mathrm{F}$ \quad  
 &
 & \quad $M^{(0\nu)}_\mathrm{GT}$ \quad & \quad $M^{(0\nu)}_\mathrm{F}$ \quad
\\[2pt]
\hline \\[-12pt]
     None    & 0.776 & $-$0.216 & & 0.997 & $-$0.304 \\
 CD-Bonn & 0.809 & $-$0.233 & & 1.045 & $-$0.327 \\
 Argonne & 0.743 & $-$0.213 & & 0.953 & $-$0.300 \\
\hline
\end{tabular}
\end{table}

\subsection{Method with the help of single-particle energy}
\label{sec:sp_energy}
Another simple way for comparing the shell-model and QRPA calculations is to identify the largest 1p1h energy of the truncated valence single-particle space with $E_\mathrm{exc}$ of Fig.~\ref{fig:rs0vbbGTF}. The single-particle energies of $^{48}$Ca are shown in Table \ref{tab:spe}, which was obtained by an updated Woods-Saxon potential \cite{Sch07}. We use this potential because it is a very realistic single-particle potential. The largest 1p1h energy in the neutron \textit{pf} valence shell is 8.8 MeV ($1f_{5/2}$-$1f_{7/2}$), and that of the \textit{sdpf} valence shell is 14.4 MeV ($1f_{5/2}$-$1d_{5/2}$). The corresponding largest 1p1h energies of the proton are 4.71 MeV (2\textit{p}$_{1/2}$-1\textit{f}$_{7/2}$) and 16.83 MeV (2\textit{p}$_{1/2}$-1\textit{d}$_{5/2}$), respectively. The largest 1p1h energy in the neutron $pf$ valence shell is close to $E_\mathrm{exc}$ = 7.5 MeV of the $pf$ valence-shell calculation discussed above. Thus, the two methods to derive the effective $E_\mathrm{exc}$ of the shell model are consistent approximately. We assume that the single-particle energies are not different appreciably for $^{48}$Ca and $^{48}$Sc. 

The $M^{(0\nu)}_\mathrm{GT}$ of the shell model are summarized in Table \ref{tab:0vbbGTSM_ratio}, and $M^{(0\nu)}_\mathrm{GT}$ of the QRPA at $E_\mathrm{exc}$ corresponding to the largest 1p1h energies of the truncated valence single-particle space are summarized in Table \ref{tab:0vbbGTQRPA_ratio}. In the application of the \textit{pf} valence shell model to $^{48}$Ca, the role of the protons is either small or none. 
Thus, we refer to the neutron excitation energy for comparing the shell-model and the QRPA calculations. The ratio of the two $M^{(0\nu)}_\mathrm{GT}$ ($sdpf$ to $pf$) of the QRPA, 1.31,  shown in Table \ref{tab:0vbbGTQRPA_ratio} is consistent with the corresponding ratio of shell model, 1.30, shown in Table \ref{tab:0vbbGTSM_ratio}. It is implied that the relative energy dependence of the mpmh effects beyond the 1p1h is close to that of the 1p1h effects. In addition, the components of the NME of the QRPA order should be included in the very-large valence shell model as mentioned before. These two points are the justification of our approach. 

\begin{table}[t]
\caption{\label{tab:spe} \baselineskip=13pt Neutron (left) and proton (right) single-particle energies associated with $^{48}$Ca obtained from the Woods-Saxon potential \cite{Sch07}. }
\centering
\begin{tabular}{ cc }
\hline \\[-13pt]
$\ $Orbital \quad & \quad Single-particle \quad \\
 & \quad energy (MeV) \\[1pt]
\hline\\[-13pt]
\multicolumn{2}{c}{ Neutron hole }  \\[1pt]
\hline\\[-11pt]
 1\textit{d}$_{5/2}$ & $-$15.61 \\
 2\textit{s}$_{1/2}$ & $-$12.55 \\
 1\textit{d}$_{3/2}$ & $-$12.53 \\
 1\textit{f}$_{7/2}$  & $-$10.00 \\[2pt]
\hline\\[-13pt]
\multicolumn{2}{c}{Neutron particle} \\[1pt]
\hline\\[-11pt]
2\textit{p}$_{3/2}$ & $-$4.60 \\
2\textit{p}$_{1/2}$ & $-$2.86 \\
1\textit{f}$_{5/2}$  & $-$1.20 \\
 1\textit{g}$_{9/2}$ & $\ \ \ \;$0.130 \\[2pt]
\hline
\end{tabular}
\quad
\begin{tabular}{ cc }
\hline \\[-13pt]
$\ $Orbital \quad & \quad Single-particle \quad \\
 & \quad energy (MeV) \\[1pt]
\hline\\[-13pt]
\multicolumn{2}{c}{ Proton hole }  \\[1pt]
\hline\\[-12pt]
 1\textit{d}$_{5/2}$ & $-$21.47 \\
 1\textit{d}$_{3/2}$ & $-$16.18 \\
 2\textit{s}$_{1/2}$ & $-$16.10 \\[2pt]
\hline\\[-14pt]
\multicolumn{2}{c}{Proton particle} \\[1pt]
\hline\\[-12pt]
1\textit{f}$_{7/2}$ & $-$9.35 \\
2\textit{p}$_{3/2}$ & $-$6.44 \\
2\textit{p}$_{1/2}$ & $-$4.64 \\[2pt]
\hline
\end{tabular}
\end{table}

\begin{table}[]
\caption{\label{tab:0vbbGTSM_ratio} \baselineskip=13pt $M^{(0\nu)}_\mathrm{GT}$ of the shell model with two different valence single-particle spaces and the ratio of $M^{(0\nu)}_\mathrm{GT}$ to that of the \textit{pf} valence-shell calculation.  }
\centering
\begin{tabular}{ ccc }
\hline \\[-11pt]
$\ \;$\multirow{2}{*}{Shell} \quad & \quad $M^{(0\nu)}_\mathrm{GT}$, \quad & \quad Ratio to  \quad \\ 
  & \quad shell model \quad & \quad $M^{(0\nu)}_\mathrm{GT} (pf)$ \quad 
\\[3pt]
\hline \\[-11pt]
    \textit{pf}    &  0.77  &   1.0 \\
    \textit{sdpf} &  1.00  & $\ \,$1.30 \\[2pt]
\hline
\end{tabular}
\vspace{10pt}
\end{table}
%
\begin{table}[]
\caption{\label{tab:0vbbGTQRPA_ratio} \baselineskip=13pt $M^{(0\nu)}_\mathrm{GT}$ of the QRPA at $E_\mathrm{exc}$ corresponding to the largest 1p1h energy of the truncated valence single-particle space (Max $E_\mathrm{exc}$). The ratio of $M^{(0\nu)}_\mathrm{GT}$ to that of the \textit{pf} valence-shell equivalent calculation is also shown. The last row shows the converged result. }
\centering
\begin{tabular}{ cccc }
\hline \\[-11pt]
 $\ $Equivalent \quad & \quad Max $E_\mathrm{exc}$ (MeV), \quad & \quad $M^{(0\nu)}_\mathrm{GT}$ of \quad & \quad Ratio to  $M^{(0\nu)}_\mathrm{GT}$ \quad \\ 
 shell  & \quad neutrons \quad & \quad QRPA \quad & \quad (\textit{pf}) of QRPA
\\[3pt]
\hline \\[-12pt]
    \textit{pf}    &   $\ \,$8.8         &  1.15 & 1.0$\ \,$ \\
    \textit{sdpf} &  14.4         &  1.51 & 1.31 \\
                      &  $\sim$ 70 &  1.88 & 1.63 \\[1pt]
\hline
\end{tabular}
\end{table}

Now, we can estimate the converged value of $M^{(0\nu)}_\mathrm{GT}$ of the shell model two ways (see Tables \ref{tab:0vbbGTSM_ratio} and \ref{tab:0vbbGTQRPA_ratio}) as 
\begin{eqnarray}
&&0.77\cdot(1.88/1.15) = 1.26, \mathrm{extension\ from\ the\ \textit{pf}\ shell}, \\[3pt]
&&1.00\cdot(1.88/1.51) = 1.25, \mathrm{extension\ from\ the\ \textit{sdpf}\ shell}.
\end{eqnarray}
These estimated values and that of Eq.~(\ref{eq:est0vbbGTSM_strfn}) are distributed in a narrow region. The result of this second method is summarized in Table \ref{tab:0vbbGTSM_est}. The estimated converged value is 64 \% larger than the \textit{pf} valence-shell value and 25 \% larger than the \textit{sdpf} valence-shell value. 

\begin{table}[]
\caption{\label{tab:0vbbGTSM_est} \baselineskip=13pt Original $M^{(0\nu)}_\mathrm{GT}$ of the shell model and estimated values corresponding to a very large valence single-particle space. }
\centering
\begin{tabular}{ ccc }
\hline \\[-11pt]
$\ $Original \quad & \quad Original $M^{(0\nu)}_\mathrm{GT}$, \quad & \quad Estimate of   \quad \\ 
  shell & \quad shell model \quad & \quad converged $M^{(0\nu)}_\mathrm{GT}$ \quad \\[3pt]
\hline \\[-12pt]
    \textit{pf}    &  0.77  &  1.26  \\
    \textit{sdpf} &  1.00  &  1.25  \\[2pt]
\hline
\end{tabular}
\end{table}

\section{ Modification of GT component of $\bm{0\nu\beta\beta}$ NME of QRPA }
\label{sec:GT0vbbNME_QRPA}
Now we estimate the mpmh effects modifying the QRPA NME. In this section, $M^{(0\nu)}_\mathrm{GT}$ indicates that of the QRPA. In principle, the best way is to compare experimental data involving the charge-change transition density and the result of the QRPA, and the possible candidate is the data of the charge-change strength functions of $1^+$ states in Fig.~\ref{fig:strfn48Ca48TiIwata}. This scheme is, however, not straightforward because the transition operator causing those data of the strength functions is not known clearly. This complexity is not surprising because the charge-exchange  reaction occurs by the nuclear force. As mentioned before, this transition operator has the IVSM component. The mixing ratio of the GT and IVSM operators is not known a priori. 

In this paper we use a simple method to evaluate the mpmh effect. That is to refer to the quenching factor already known for the strength function of the shell model with the GT operator so as to reproduce the data. In Ref.~\cite{Iwa15} the quenching factor of 0.77 is applied to the GT operator for both \textit{pf} and \textit{sdpf} valence-shell calculations as mentioned before. On the other hand, the quenching factor of the QRPA calculation for the GT strength function is 0.5 for $^{48}$Ca$\rightarrow$$^{48}$Sc and 0.38 for $^{48}$Ti$\rightarrow$$^{48}$Sc \cite{Ter18}. These factors were determined in the low-energy regions close to those discussed for the shell model. There are two origins of the quenching factor to the QRPA transition strength. One is the effect of the mpmh correlations missing in the QRPA nuclear wave functions, and another is the modification of the transition operator, e.g., the IVSM components and the vertex corrections. 
It is known that the mpmh correlations have effects to reduce the charge-change transition strength in low-energy regions \cite{Ber82,Dan97,Rob19} and the NME \cite{Cau08}. 
The shell model needs only the quenching factor due to the operator-modification origin. The quenching factor to the QRPA is the product of the factors due to the two origins, and the one due to the operator-modification origin is shared by the two methods. For $^{48}$Ca this idea leads to a relation of the quenching factors to the strength functions 
\begin{eqnarray}
0.5 = 0.77^2 x,
\end{eqnarray}
where $x$ is the quenching factor due to the mpmh correlations of the nuclear wave functions, and for $^{48}$Ti
\begin{eqnarray}
0.38 = 0.77^2 x^\prime,
\end{eqnarray} 
where $x^\prime$ is the same as $x$ but for $^{48}$Ti. Thus, we obtain $x$ = 0.847 and $x^\prime$ = 0.644. The difference between these two values indicates that the mpmh effect is larger in $^{48}$Ti than in $^{48}$Ca. The QRPA is a reasonable approximation to the doubly magic $^{48}$Ca, however, the quality of approximation is deteriorated for $^{48}$Ti. Thus, the $0\nu\beta\beta$ NME is affected. Note that the quenching factors are applied in the low-energy regions where the shell model is reliable. 

We introduce four modification factors to $\langle F| c_p^\dagger c_n |B_F\rangle$ and $\langle B_I | c_{p\prime}^\dagger c_{n^\prime} | I \rangle$ when they are used with the GT operator [see Eq.~(\ref{eq:m0vbbGTQRPA})]:
\begin{itemize}
\item the quenching factor $q^F_l$ multiplied to $\langle F| c_p^\dagger c_n |B_F\rangle$ with $|B_F\rangle$ in the low-energy region corresponding to the reliable region of the shell model, 
\item the enhancing factor $q^F_h$ multiplied to the same transition-density matrix element but for the higher-energy region, and 
\item $q^I_l$ and $q^I_h$ which are the same as $q^F_l$ and $q^F_h$ but for $\langle B_I | c_{p\prime}^\dagger c_{n^\prime} | I \rangle$, respectively. 
\end{itemize}
The enhancing factors are introduced for the higher-energy region because of the sum rule of the transition strength. The modification of the transition-density matrix is shared by the strength function associated with the charge-exchange reaction and the NME of the double-$\beta$ decay involving the spin transition operator. 

The partition of the energy region and the quenching factor in the lower-energy region were determined on the basis of the smoothened strength functions. Therefore the mpmh effects on the $0\nu\beta\beta$ NME should be simulated in the same manner. With those modification factors, smoothening the intermediate-state energy dependence by the Lorentzian function, and the truncation up to the first order with respect to $(1-q)$ ($q$ is any of the four modification factors), the following equation of the modified $M^{(0\nu)}_\mathrm{GT}$ is obtained: 
\begin{eqnarray}
M^{(0\nu)}_\mathrm{GT}(\mathrm{modified}) &=& M^{(0\nu)}_\mathrm{GT} + M^{(0\nu)}_{\mathrm{GT}\textrm{-}1l} +
M^{(0\nu)}_{\mathrm{GT}\textrm{-}1h}, \label{eq:m0vbbGTQRPA_mdfd}
\end{eqnarray}
\begin{eqnarray}
M^{(0\nu)}_{\mathrm{GT}\textrm{-}1l} &=& 
-(1-q^F_l) \frac{1}{\pi}\int_{-\infty}^{E^F_c} dE 
\sum_{E_{BF}} \frac{\varepsilon P_F(E_{BF})}{(E_{BF}-E)^2+\varepsilon^2}
\nonumber \\
&&-(1-q^I_l) \frac{1}{\pi}\int_{-\infty}^{E^I_c} dE 
\sum_{E_{BI}} \frac{\varepsilon P_I(E_{BI})}{(E_{BI}-E)^2+\varepsilon^2}\,, 
\label{eq:m0vbbGTQRPA_1l}
\end{eqnarray}
\begin{eqnarray}
M^{(0\nu)}_{\mathrm{GT}\textrm{-}1h} &=& 
-(1-q^F_h) \frac{1}{\pi}\int_{E^F_c}^{\infty} dE 
\sum_{E_{BF}} \frac{\varepsilon P_F(E_{BF})}{(E_{BF}-E)^2+\varepsilon^2}
\nonumber \\
&&-(1-q^I_h) \frac{1}{\pi}\int_{E^I_c}^{\infty} dE 
\sum_{E_{BI}} \frac{\varepsilon P_I(E_{BI})}{(E_{BI}-E)^2+\varepsilon^2}\,, 
\label{eq:m0vbbGTQRPA_1h}
\end{eqnarray}
\begin{eqnarray}
P_F(E_{BF}) &=& \sum_{B_I} \sum_{pn p^\prime n^\prime}
\langle pp^\prime| V^{(0\nu)}_\mathrm{GT}(r;{\bar{E}}_B) |nn^\prime\rangle
\langle F| c^\dagger _p c_n | B_F \rangle  
\langle B_F | B_I \rangle 
\langle B_I | c^\dagger_{p^\prime} c_{n^\prime} | I \rangle, \label{eq:PF}
\end{eqnarray}
\begin{eqnarray}
P_I(E_{BI}) &=& \sum_{B_F} \sum_{pn p^\prime n^\prime}
\langle pp^\prime| V^{(0\nu)}_\mathrm{GT}(r;{\bar{E}}_B) |nn^\prime\rangle
\langle F| c^\dagger _p c_n | B_F \rangle \langle B_F | B_I \rangle 
\langle B_I | c^\dagger_{p^\prime} c_{n^\prime} | I \rangle , \label{eq:PI}
\end{eqnarray}
where $E_{BI}$ and $E_{BF}$ are the QRPA eigen energies of $|B_I\rangle$ and $|B_F\rangle$, respectively. 
The parameter $\varepsilon$ $>$ 0 is a constant chosen so as to reproduce the width of the experimental strength function. $E^I_c$ and $E^F_c$ are the QRPA energies distinguishing the lower- and higher- energy regions associated with the initial and final states, respectively. The lowest QRPA energy of the transition to the intermediate nucleus is identified with the transition energy to the ground state of the intermediate nucleus. $E^I_c$ and $E^F_c$ are obtained by adding these lowest QRPA energies to the boundary excitation energies discussed above. $E^I_c$ = 12.053 MeV and $E^F_c$ = 17.387 MeV are obtained from our numerical calculation. The deviation of the modification factors from one represents the modification effects. Thus, if there is no modification, the modification terms vanish. Since the quenching factors are less than one, these factors decrease the NME in the lower-energy region. The enhancing factors increase the NME in the higher-energy region. 

We use
\begin{eqnarray}
&&q^I_l = \sqrt{x} = 0.92, \mathrm{and} \\
&&q^F_l = \sqrt{x^\prime} = 0.80.
\end{eqnarray}
The transition strength up to $E_\mathrm{exc}$ = 13 MeV is 19.575 for $^{48}$Ca $\rightarrow$ $^{48}$Sc (the integral of the QRPA strength function). The redundant transition strength in $E_\mathrm{exc}$ $<$ 13 MeV due to the insufficiency of the mpmh correlations is estimated 
\begin{eqnarray}
19.575 (1-x) = 2.99. 
\end{eqnarray}
The transition strength in the higher-energy region is 4.96. The enhancing factor is calculated
\begin{eqnarray} 
q^I_h = \sqrt{(2.99 + 4.96) / 4.96} = 1.27. 
\end{eqnarray}
In the same manner, we obtain $q^F_h$ = 1.44 for $^{48}$Ti.

We discuss the uncertainty of our approach and a further modification of the equation for the QRPA NME. The mpmh correlations have two effects. One is to modify the QRPA states, and another is to create mpmh states not included in the QRPA. When the GT strength in the low-energy region is reduced by the mpmh correlations, the absolute values of the major transition-density matrix elements of the QRPA states have to decrease sufficiently overwhelming the effect of the mpmh states to increase the strength. Thus, the reduction of the contribution of the QRPA states to the NME is the main effect of the modification in the lower-energy region. The sum rule implies that the shift of the strength of the transition-density matrix elements occurs to the higher-energy region. This shift enhances the contribution of the QRPA states to the $0\nu\beta\beta$-decay NME in that region, if the change occurs uniformly to those matrix elements. Equation (\ref{eq:m0vbbGTQRPA_mdfd}) expresses this effect. The effect of the mpmh states is not negligible in the higher-energy region. In the continuum region, it is known as the spreading width, e.g., \cite{Har01}. In the reality, the importance of the mpmh-state effects increases gradually as the excitation energy increases. We introduced a simplification that this effect is significant only in the higher-energy region. The contribution of the mpmh states to the NME of the double-$\beta$ decay cannot be estimated by our approach because this specific effect may reduce or enhance the NME. The sign of the contribution is unknown without the calculations including the mpmh effects in the higher-energy regions. 

Equation (\ref{eq:m0vbbGTQRPA_mdfd}) expresses an extreme case that the mpmh corrections in the higher-energy region arise entirely through the modification of the QRPA states; we call this case extreme case I. It is possible to consider another extreme case that the mpmh corrections in the higher-energy region are entirely carried by the new states beyond the QRPA; we call this case extreme case II. Three  sub-extreme cases belonging to extreme case II can be discussed as follows. 

\begin{figure}[t]
\centering
\includegraphics[width=0.5\columnwidth]{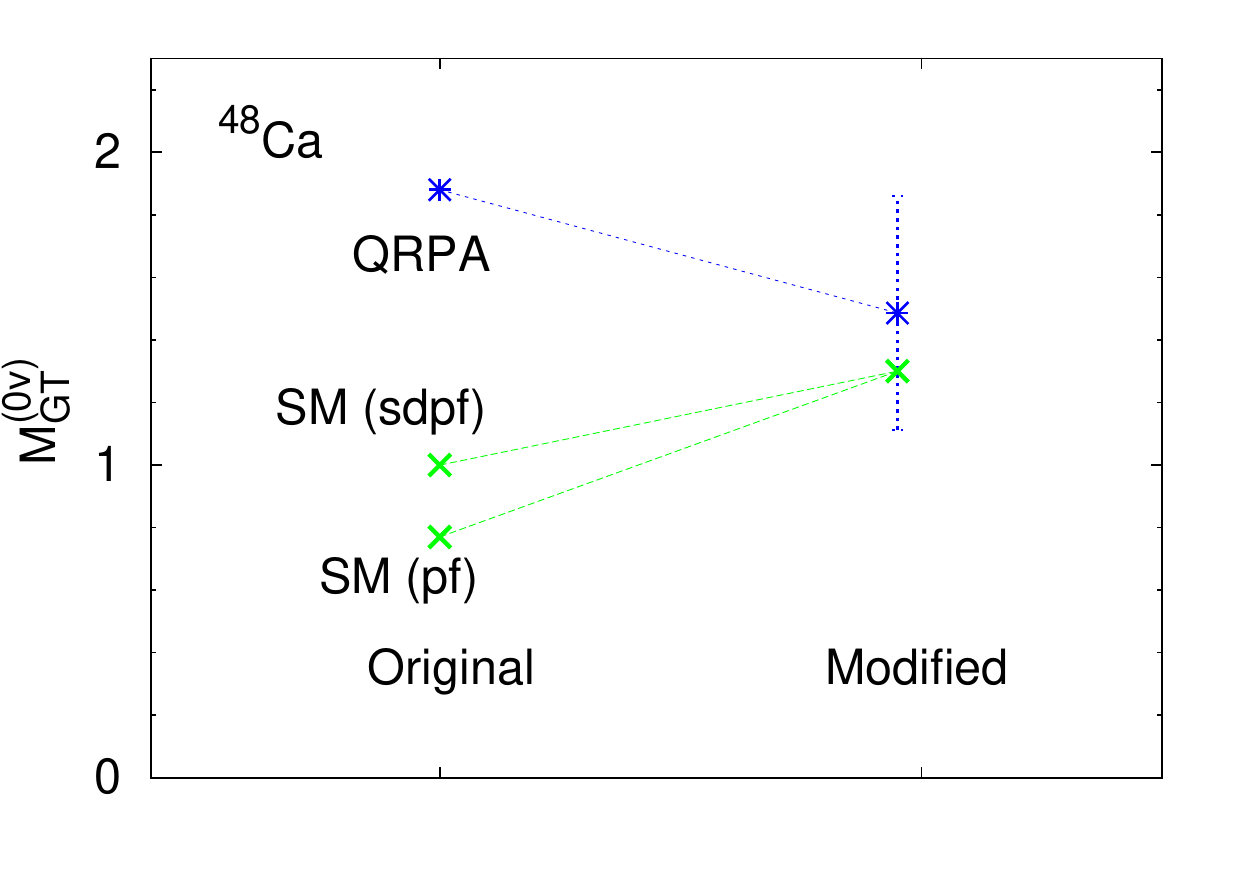}
\vspace{-10pt}
\caption{ \protect \label{fig:0vbbGT_m} \baselineskip=13pt The values of $M_\mathrm{GT}^{(0\nu)}$ of the shell model and QRPA labeled ``Original'' and  the modified $M_\mathrm{GT}^{(0\nu)}$ labeled ``Modified''. The shell model, denoted by SM, and the QRPA are distinguished by the different symbols. The modified shell-model value is the average of 1.25, 1.26, and 1.27 obtained in Sec.~\ref{sec:GT0vbbNME_SM}. The error bar of the modified QRPA shows the uncertainty range, and the mean value is interpreted as the value in the randomness limit; see text.  }
\end{figure}

\begin{enumerate}
\item[i. ]
The mpmh-state corrections are maximally coherent to $M^{(0\nu)}_\mathrm{GT}$. Usually, the effects of the mpmh states are smaller than those of the 1p1h states, e.g., for the transition strength. Thus, $M^{(0\nu)}_\mathrm{GT}$ in this extreme case may not exceed the $M^{(0\nu)}_\mathrm{GT}$(modified) of Eq.~(\ref{eq:m0vbbGTQRPA_mdfd}) derived in extreme case I. Thus, Eq.~(\ref{eq:m0vbbGTQRPA_mdfd}) also gives the upper limit of extreme case II. 

\item[ii. ]
 The mpmh-state corrections are maximally anticoherent to $M^{(0\nu)}_\mathrm{GT}$. In this case the lower limit of  $M^{(0\nu)}_\mathrm{GT}$(modified) is estimated  
\begin{eqnarray}
M^{(0\nu)}_\mathrm{GT} + M^{(0\nu)}_{\mathrm{GT}\textrm{-}1l} -
M^{(0\nu)}_{\mathrm{GT}\textrm{-}1h}. \label{eq:m0vbbGTQRPA_mdfd_ll}
\end{eqnarray}

\item[iii. ]
The mpmh-state corrections are accompanied by the strong randomness, and the corrections are canceled. In this extreme case $M^{(0\nu)}_\mathrm{GT}$(modified) is given by 
\begin{eqnarray}
M^{(0\nu)}_\mathrm{GT} + M^{(0\nu)}_{\mathrm{GT}\textrm{-}1l}\,.
\end{eqnarray}
\end{enumerate}
Thus, we can evaluate the uncertainty range of $M^{(0\nu)}_\mathrm{GT}$(modified) as 
\begin{eqnarray}
M^{(0\nu)}_\mathrm{GT} + M^{(0\nu)}_{\mathrm{GT}\textrm{-}1l} -
M^{(0\nu)}_{\mathrm{GT}\textrm{-}1h} 
< M^{(0\nu)}_\mathrm{GT}(\mathrm{modified})
< M^{(0\nu)}_\mathrm{GT} + M^{(0\nu)}_{\mathrm{GT}\textrm{-}1l} +
M^{(0\nu)}_{\mathrm{GT}\textrm{-}1h} \,.
\label{eq:m0vbbGTQRPA_mdfd_uncn} 
\end{eqnarray}
The $M^{(0\nu)}_\mathrm{GT}$(modified) of extreme case I is equal to the upper limit of extreme case II. It is speculated that the upper limit in a mixing case between the two extreme cases does not exceed significantly the upper limit of Eq.~(\ref{eq:m0vbbGTQRPA_mdfd_uncn}). If one effect is weakened, and another effect appears, the total may not change significantly. Total effect to the higher-energy region is restricted because the change in the transition density is restricted by the sum rule. A similar discussion is possible for the lower limit in the mixing case. 
$M_{\mathrm{GT}\textrm{\small -}1l}^{(0\nu)}$ = $‒$0.394 and 
$M_{\mathrm{GT}\textrm{\small -}1h}^{(0\nu)}$ = 0.374 are obtained numerically, and consequently the uncertainty range is obtained
\begin{eqnarray}
1.112 < M^{(0\nu)}_\mathrm{GT}\mathrm{(modified)} < 1.860,
\end{eqnarray}
by referring to the value of  $M_\mathrm{GT}^{(0\nu)}$ of the QRPA in Sec.~\ref{sec:GT0vbbNME_SM}. The value of $M_\mathrm{GT}^{(0\nu)}$(modified) in the limit of the randomness of case iii is 1.486. Our result is summarized in Fig.~\ref{fig:0vbbGT_m}. The modified QRPA value in the randomness limit is much closer to the converged value by our estimation of the shell model than the original one. The mpmh correlations are generally accompanied by the randomness in high-energy regions. This tendency is indicated by, e.g., the success of the random-matrix theory, e.g.,  \cite{Boh69,Sev18}. Therefore, we speculate that the best modified value of the QRPA is closer to the randomness limit than the edge values of the coherent limit. Our result shows that the consistency of the shell model and the QRPA is possible to obtain by taking into account the sufficiently large valence single-particle space and the mpmh correlations. 

\section{ Fermi component of $\bm{0\nu\beta\beta}$ NME }
\label{sec:F0vbbNME}
According to Ref.~\cite{Iwa16}, the $M^{(0\nu)}_\mathrm{F}$ of the \textit{pf} valence-shell calculation averaged for the two SRC methods is $‒$0.223, and that of the \textit{sdpf} valence-shell calculation is $‒$0.314, where no quenching factor is used. The latter is 41 \% larger than the former in absolute value.  The corresponding increasing ratio of the QRPA read from Fig.~\ref{fig:rs0vbbGTF} is 21 \%; the $M^{(0\nu)}_\mathrm{F}$ value is $‒$0.19 at $E_\mathrm{exc}$ = 8.8 MeV corresponding to the \textit{pf} valence shell, and $‒$0.23 at $E_\mathrm{exc}$ = 14.4 MeV corresponding to the \textit{sdpf} valence shell. These equivalent $E_\mathrm{exc}$'s were used in Table \ref{tab:0vbbGTQRPA_ratio}. There is a non-negligible difference in the increasing ratio of the two methods. The converged value of $M^{(0\nu)}_\mathrm{F}$ of the QRPA is $‒$0.35, of which the absolute value is increased from the value at 14.4 MeV by 52 \%. Those $M^{(0\nu)}_\mathrm{F}$ values are summarized in Table \ref{tab:0vbbFSMQRPA}. 
The absolute value of $M^{(0\nu)}_\mathrm{F}$ of the shell model is larger than that of the QRPA for the same $E_\mathrm{exc}$. This relation of magnitude is inverted in comparison to that for $M^{(0\nu)}_\mathrm{GT}$; see Tables 
\ref{tab:0vbbGTQRPA_ratio} and \ref{tab:0vbbGTSM_est}.   
According to the method that we applied to $M^{(0\nu)}_\mathrm{GT}$ in Sec.~\ref{sec:GT0vbbNME_SM}, we have two estimates of the extrapolated $M^{(0\nu)}_\mathrm{F}$ of the shell model: 
\begin{eqnarray}
&&-0.223 \cdot (0.35/0.19) = -0.411, \ \textrm{and} \nonumber \\
&&-0.314 \cdot (0.35/0.23) = -0.478, \label{eqn:0vbbFSM_est}
\end{eqnarray}
(see Table \ref{tab:0vbbFSMQRPA}). 

\begin{table}[]
\caption{\label{tab:0vbbFSMQRPA} \baselineskip=13pt $M^{(0\nu)}_\mathrm{F}$ of the shell model for the two different valence shells and those of the QRPA for the equivalent $E_\mathrm{exc}$'s and a very large $E_\mathrm{exc}$. }
\centering
\begin{tabular}{ cccc }
\hline \\[-11pt]
  Shell \quad & \quad Equivalent \quad & \quad $M^{(0\nu)}_\mathrm{F}$ (shell \quad & \quad $M^{(0\nu)}_\mathrm{F}$ \quad \\
                        & \quad max $E_\mathrm{exc}$ (MeV) & model) \quad & \quad (QRPA) \quad \\[2pt] 
\hline \\[-12pt]
    \textit{pf}    &  $\ \,$8.8 &  $-$0.223 &  $-$0.19   \\
    \textit{sdpf} &    14.4  &  $-$0.314 &  $-$0.23 \\
  $\ $Large-space limit           &             &                &  $-$0.35 \\[1pt]
\hline
\end{tabular}
\end{table}

For considering $M^{(0\nu)}_\mathrm{F}$ of the QRPA we discuss the transition-operator dependence of the modification factor from a general viewpoint. To our knowledge, this dependence can be summarized as follows:
\begin{itemize}
\item
the appreciable quenching is necessary for the GT transition whether it is caused by the strong or weak interaction at least for the transitions with not-high transition energies corresponding to a one-major valence shell (see Sec.~\ref{sec:GT0vbbNME_QRPA}) or the spin-orbit splitting; this is common for the QRPA and shell model. The appropriate quenching factor depends on the method. 
\item
The isobaric-analog transition does not need a quenching factor at least for the QRPA. This is not surprising because the Fermi transition strength concentrates on the isobaric-analog state, e.g., \cite{Pup11}, and the QRPA satisfies the Fermi sum rule. 
\item
For the electric transitions, the necessity of the effective charge is much lower than that of the quenching factor for the GT operator except for keeping the center of mass, e.g., \cite{Ter08}. This is particularly clear for the nuclei to which the QRPA is a good approximation. 
\item
The magnetic transition needs the effective $g$ factor, e.g., \cite{Tow87,Hey90}. The adjustment of the spin $g$  factor is significant. 
\end{itemize}
When these features are stated for the QRPA, sufficiently large single-particle spaces are assumed. From this observation, we obtain an idea that the transition operator is essential for the modification factors rather than the difference of the interactions, that is, 
\begin{itemize}
\item
the spin operator has the high necessity of the quenching factor at least for the transitions with the not-high energies. 
\item
Coordinate operators do not have that necessity at least for the nuclei to which the QRPA is a good approximation. 
\item
The operator only changing the charge does not have that necessity. 
\end{itemize}
Based on the last two items, we do not quench $M^{(0\nu)}_\mathrm{F}$ of the QRPA. 

\section{\label{sec:0vbbNME}Estimated $\bm{0\nu\beta\beta}$ NME}
$M^{(0\nu)}$ can be calculated using the modified components. We did not pinpoint the modified components of $M^{(0\nu)}$ of the QRPA due to the uncertainty of our estimation. In this section, we make a choice for simplifying the comparison. We use the QRPA $M^{(0\nu)}_\mathrm{GT}$(modified) obtained with the complete randomness of the mpmh correlations in the high-energy region on the basis of the discussion in Sec.~\ref{sec:GT0vbbNME_QRPA}. For the Fermi component of the shell model, we do not have a speculation on what value is more likely than others in the range of $-M^{(0\nu)}_\mathrm{F}$ = 0.411$-$0.478. The modified components of the shell-model and QRPA NMEs are summarized in Table \ref{tab:0vbbGTFSMQRPA_est}.

\begin{table}[]
\caption{\label{tab:0vbbGTFSMQRPA_est} \baselineskip=13pt Estimated $M^{(0\nu)}_\mathrm{GT}$ and $M^{(0\nu)}_\mathrm{F}$ from the shell model and the QRPA. }
\centering
\begin{tabular}{ ccc }
\hline \\[-12pt]
$\ $Component  \quad & \quad Estimate from \quad & \quad  Estimate from \quad  \\
                                 & shell model  &  QRPA \quad \\[1pt] 
\hline \\[-8pt]
   $M^{(0\nu)}_\mathrm{GT}$   &  1.26      &  1.49 \\[2pt]
   $-M^{(0\nu)}_\mathrm{F}$    &     0.41$-$0.48 &  0.35    \\[1pt]
\hline
\end{tabular}
\end{table}

By using $g_V$ = 1.0, and tentatively the bare value of $g_A$ = 1.276 \cite{Mar19}, the estimated $M^{(0\nu)}$ based on the shell model for $^{48}$Ca is found to be 1.512$-$1.554, 
and that from the QRPA is 1.705. 
A partial cancellation occurs between the difference in $M^{(0\nu)}_\mathrm{GT}$ of the two methods and that in $M^{(0\nu)}_\mathrm{F}$ of these  methods as seen from Table \ref{tab:0vbbGTFSMQRPA_est}. The values of 1.5$‒$1.7 are situated in the middle of the distribution of $M^{(0\nu)}$ by various methods; see, e.g., Ref.~\cite{Eng17}, in which the $M^{(0\nu)}$'s are in the range of 0.6 to 3.0 for $^{48}$Ca. 

\section{ NME of $\bm{2\nu\beta\beta}$ decay }
\label{sec:2vbbNME}
Next physical quantities to examine are the components of the $2\nu\beta\beta$ NME $M^{(2\nu)}_\mathrm{GT}$ and $M^{(2\nu)}_\mathrm{F}$; the former is the GT component, and the latter is the Fermi component. 
The former is defined by 
\begin{eqnarray}
M^{(2\nu)}_\mathrm{GT} = 
\sum_B \frac{1}{E_B - \bar{M}} \langle F | \bm{\sigma} \tau^- | B\rangle 
\cdot \langle B | \bm{\sigma} \tau^- | I \rangle, \label{eq:m2vbbGT}
\end{eqnarray}
where $\bar{M}$ denotes the mean value of the masses of the initial and final nuclei. 
The closure approximation is not applied to the NME of the $2\nu\beta\beta$ decay. 
$M^{(2\nu)}_\mathrm{GT}$ of the QRPA is written using the two sets of the intermediate states as  
\begin{eqnarray}
M^{(2\nu)}_\mathrm{GT}(\mathrm{QRPA}) = 
\sum_{B_I B_F} \frac{1}{E_B - \bar{M}} \langle F | \bm{\sigma} \tau^- | B_F \rangle
\cdot\langle B_F | B_I \rangle 
\langle B_I | \bm{\sigma} \tau^- | I \rangle. \label{eq:m2vbbGTQRPA}
\end{eqnarray}
$E_B$ is the energy of the intermediate state. We use $E_{BI}$ with an energy calibration for $E_B$ because the QRPA is better for $^{48}$Ca than $^{48}$Ti. 
The equation of $M^{(2\nu)}_\mathrm{F}$ can be obtained by removing the spin operators in Eqs.~(\ref{eq:m2vbbGT}) and (\ref{eq:m2vbbGTQRPA}). The intermediate states contributing to $M^{(2\nu)}_\mathrm{GT}$ and $M^{(2\nu)}_\mathrm{F}$ are different. 
The $2\nu\beta\beta$ NME is calculated by   
\begin{eqnarray}
M^{(2\nu)} = M^{(2\nu)}_\mathrm{GT} - \left(\frac{g_V}{g_A}\right)^2 M^{(2\nu)}_\mathrm{F}.  
\end{eqnarray}

\begin{figure}[]
\centering
\includegraphics[width=0.5\columnwidth]{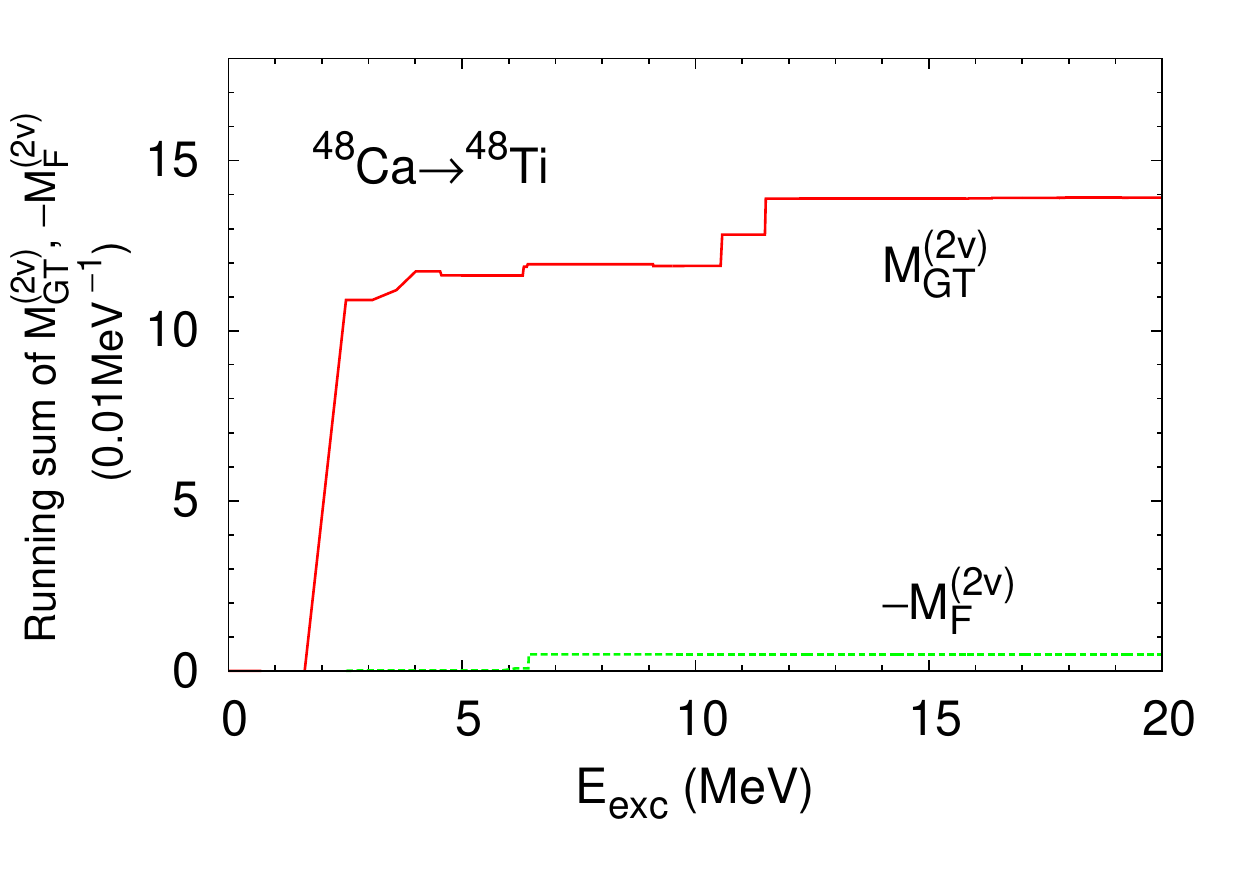}
\vspace{-10pt}
\caption{ \protect \label{fig:rs2vbbGTF} \baselineskip=13pt $M^{(2\nu)}_\mathrm{GT}$ and $M^{(2\nu)}_\mathrm{F}$ of the QRPA as functions of $E_\mathrm{exc}$ of the intermediate nucleus $^{48}$Sc.  $E_\mathrm{exc}$ is obtained from the QRPA energy of the calculation based on the initial state with a calibration energy adjusted to the $1^+$ state of the lowest-energy peak of the experimental GT  strength function. No quenching factor is  used. }
\end{figure}

The running sums of these components of the QRPA are shown in Fig.~\ref{fig:rs2vbbGTF}. The horizontal axis indicates again the excitation energy of the intermediate state. $M^{(2\nu)}_\mathrm{GT}$ and $M^{(2\nu)}_\mathrm{F}$ of the QRPA are 0.139 MeV$^{-1}$ and $-$0.0048 MeV$^{-1}$, respectively. $M^{(2\nu)}_\mathrm{F}$ is very small compared to $M^{(2\nu)}_\mathrm{GT}$ because of the approximate isospin invariance. The lowest-energy contributions occupy nearly 85 \% of $M^{(2\nu)}_\mathrm{GT}$, and the second largest contribution around $E_\mathrm{exc}$ = 11 MeV is of the giant resonance of $^{48}$Ca$\rightarrow$$^{48}$Sc; see Fig.~\ref{fig:strfn48Ca48TiIwata}. For $M^{(2\nu)}_\mathrm{F}$, almost 100 \% is occupied by the lowest-energy contribution. The running sum of Fig.~\ref{fig:rs2vbbGTF} indicates that the relatively small valence single-particle space is sufficient for the convergence of $M^{(2\nu)}_\mathrm{GT}$ and $M^{(2\nu)}_\mathrm{F}$. $M^{(2\nu)}$ of 0.142 MeV$^{-1}$ is obtained with the bare values of $g_A$ = 1.276 and $g_V$ = 1.0. This is a result with no quenching factor. 

The authors of Ref.~\cite{Iwa16} derived the quenching factor to the GT operator of 0.74 (the  \textit{pf} valence shell) and 0.71 (the \textit{sdpf} valence shell) from the experimental data of the GT$^+$ and GT$^-$ strengths. They calculated $M^{(2\nu)}$ with these quenching factors and obtained 0.052 MeV$^{-1}$ (the \textit{pf} valence shell, GXPF1B interaction) and 0.051 MeV$^{-1}$ (the \textit{sdpf} valence shell, SDPFMU-DB interaction). The closeness of these two values also indicates the sufficiency of the relatively small valence single-particle space. Their results are close to the early experimental value of 0.046 $\pm$ 0.004 MeV$^{-1}$ \cite{Bara15}. It is noted, however, that the experimental half-life of $^{48}$Ca to the $2\nu\beta\beta$ decay was updated recently \cite{Bar19} and increased in comparison to the previous one.  By using the new data and the method used in Ref.~\cite{Iwa16}, the revised experimental $M^{(2\nu)}$ of 0.042 $\pm$ 0.004 MeV$^{-1}$ is obtained.  

Let us adjust $M^{(2\nu)}$ of the QRPA for comparison using the manner of Ref.~\cite{Iwa16}. We apply the quenching factors of 0.74 and 0.71 of the GT operator to the QRPA result; $M^{(2\nu)}_\mathrm{GT}$ of the QRPA is reduced by factors of 0.55 and 0.50, respectively. The $M^{(2\nu)}$ of the QRPA becomes  0.0745 MeV$^{-1}$ (the average of the results with the different quenching factors), which is 45 \% larger than the shell-model value. It is stressed that this difference is much smaller than the corresponding difference in $M^{(0\nu)}$, which was nearly a factor of two;   see Fig.~\ref{fig:0vbbGT_m}. The reason is that the relatively small energy region is sufficient for the intermediate states, and this is because the neutrino potential is not used.

Next, we introduce our original modification method. 
The modification of $M^{(2\nu)}$ of the QRPA is similar to that of $M^{(0\nu)}$ but simpler.  
In the previous discussion for $M^{(0\nu)}_\mathrm{GT}$, the NME of the QRPA was reduced in the low-energy region by 26 \% ($1-\sqrt{xx^\prime}$) by referring to the ratios of the quenching factors of the shell model and the QRPA for simulating the mpmh effects. For $M^{(2\nu)}$, we refer to the GT$^-$ strength of the first $1^+$ state at $E_\mathrm{exc}$ = 2.5173 MeV (experimental value) \cite{nndc} because this state is exclusively important for the $2\nu\beta\beta$ decay of $^{48}$Ca. The ratio of the GT$^-$ strength of the shell model to the corresponding value of the QRPA is 0.894, which is used as an extra quenching factor to the QRPA $M^{(2\nu)}_\mathrm{GT}$ by our modification method. 
As $M^{(0\nu)}_\mathrm{F}$ discussed in  Sec.~\ref{sec:F0vbbNME}, we do not modify $M^{(2\nu)}_\mathrm{F}$. We also do not consider the modification in the higher-energy region. The modified result of $M^{(2\nu)}$ of the QRPA is 0.0666 MeV$^{-1}$ (again the average). This value is 29 \% larger than the average shell-model value of 0.0515 MeV$^{-1}$. 

The modification of $M^{(2\nu)}$ of the shell model is made due to the giant resonance. This resonance is located in the energy region where the strength function of $^{48}$Ti$\rightarrow$$^{48}$Sc calculated by the one-major valence shell model  does not have any strength; see Fig.~\ref{fig:strfn48Ca48TiIwata}. Thus, the shell-model calculation of the double-$\beta$ decay does not have the contribution of the giant resonance. Its contribution in the QRPA without the quenchings is 0.0198 MeV$^{-1}$; see Fig.~\ref{fig:rs2vbbGTF}. We add the value with the quenching corrections 
\begin{eqnarray}
 0.0198 \cdot 0.525 \cdot 0.894 \ \mathrm{MeV}^{-1} = 0.0093 \ \mathrm{MeV}^{-1}, 
\end{eqnarray}
to the shell-model value of 0.0515 MeV$^{-1}$ and obtain 0.0608 MeV$^{-1}$, which is only 9 \% smaller than the modified QRPA value of 0.0666 MeV$^{-1}$. Therefore, the consistency of the two models can be obtained by our modification method. The result of this discussion is summarized in Table \ref{tab:2vbbSMQRPA}. 

\begin{table}[]
\caption{\label{tab:2vbbSMQRPA} \baselineskip=13pt $M^{(2\nu)}$ of the shell model and the QRPA with the shell-model quenching factor multiplied (original) and those modified by our method (modified). The modification method depends on the model; see text. }
\centering
\begin{tabular}{ ccc }
\hline \\[-11pt]
   & \quad $M^{2\nu}$ of shell model \quad & \quad  $M^{2\nu}$ of QRPA \quad  \\
   & (MeV$^{-1})$ & (MeV$^{-1}$) \\[2pt]
\hline \\[-11pt]
   Original    &  0.0515 &  0.0745 \\[2pt]
   Modified   &  0.0608 &  0.0666 \\[1pt]
\hline
\end{tabular}
\end{table}

If i) the relevant energy region is low, ii)  the many-body correlations in the initial and final states are small, and iii) the configuration mixing by the transition operator is small, the two models are  approximately consistent. The first condition is obviously fulfilled for $M^{(2\nu)}$ because the neutrino potential is not used. The second condition affects the applicability of the QRPA, which depends on the nucleus. The third condition is better satisfied for $M^{(2\nu)}$ than $M^{(0\nu)}$ again because of no neutrino potential. 

\section{ Summary }
\label{sec:summary}
We have proposed a new method to estimate the NME components of the double-$\beta$ decays from the information already available and demonstrated that the similar results can be obtained for the $0\nu\beta\beta$ NME of $^{48}$Ca $\rightarrow$ $^{48}$Ti of the shell model and the QRPA with the modifications. The extrapolated $M^{(0\nu)}_\mathrm{GT}$ of the shell model is 1.26 (the average of three very close values), and the modified result of the QRPA is expressed by the range of 1.112$-$1.860. We speculate that the likely value is close to the mean value of 1.486 rather than the edge values. The difference of the two methods was a factor of two before our modifications. 

The current shell model with the one-major valence shell is not sufficient for the $0\nu\beta\beta$ decay, and the QRPA does not have as much mpmh effects as the shell model has. The result of the shell model was modified by speculating the dependence on the particle-hole energy representing the valence shells based on the intermediate-state energy dependence of the running sum of the QRPA NME. This speculation is justified by two points. One is the fact that the increasing rate of the $0\nu\beta\beta$ GT NME with respect to that representative, or intermediate state, energy is the same between the shell model and QRPA in the discussed range. Another is that the NME of the shell model with a very-large valence space should include the NME of the QRPA.  The experimental data of the charge-change strength functions were used for clarifying the reliable energy region of the shell model. The necessary energy region for the $0\nu\beta\beta$ NME is rather large because of the neutrino potential. The QRPA calculation can be performed up to the convergence of the result with respect to the single-particle as well as the intermediate-state energies. Thus, the QRPA was used for modifying the shell-model result. 

The insufficiency of the QRPA was evaluated from the experimental data of the strength function and the shell-model calculation in the lower-energy region, in which the shell model is very reliable. The mpmh effects seem more important for $^{48}$Ti than $^{48}$Ca. This analysis was applied to the modification of the QRPA result of the double-$\beta$ NMEs. This application is enabled by the transition-density matrix shared by the different phenomena. This modification method to the QRPA has the uncertainty in the higher-energy region, thus, we considered the multiple extreme cases. 

The key point of our approach is to combine the information of the different methods complementarily paying attention to the running sums of the NMEs with respect to the intermediate-state energy. In this paper, we concentrated on $^{48}$Ca$\rightarrow$$^{48}$Ti because the sufficient information for our approach is available only for this decay instance currently. That is the shell-model calculations with different valence single-particle spaces and the data of the strength functions. The GT and Fermi components were investigated separately for the $0\nu\beta\beta$ decay. The modified $0\nu\beta\beta$ NME turned out to be in the middle of the distribution of the NME of many other calculations. 

The $2\nu\beta\beta$ NME was also studied. Those NME of the shell model and the QRPA are 0.0515 MeV$^{-1}$ and 0.0745 MeV$^{-1}$, respectively. Both values were  obtained using the common quenching factors to the GT operators around 0.73, which is not the fitting parameter to the corresponding experimental value of 0.042 $\pm$ 0.004 MeV$^{-1}$. We obtained the modified shell-model result of 0.0608 MeV$^{-1}$  taking into account the contribution of the giant resonance. We also obtained the modified QRPA result of 0.0666 MeV$^{-1}$ including the extra quenching factor of 0.894 to the $2\nu\beta\beta$ NME reflecting the insufficient mpmh effects of the QRPA. If the relevant region of the intermediate-state energy is small, i.e.~there is no singularity of the transition operator, and the mpmh correlations are not significant compared to the 1p1h correlations, the consistency of the two models can be obtained by the simple modifications. 

There are two reasons why our approach is important. One is that the discrepancy problem of the $0\nu\beta\beta$ NME by the methods has been unsolved more than thirty years; we need to clarify the problems. Another is that the true values of the $0\nu\beta\beta$ NME are required whatever method is used; there is no precondition for the method to use. Thus, we can use as much information as possible. Finally, the improvement of each method, i.e., the shell model, QRPA, and others, by extending the wave functions should continue in order to confirm the correct $0\nu\beta\beta$ NME by the double convergence mentioned in the introduction eventually. 

\begin{acknowledgments}
This study is supported by European Regional Development Fund, Project ``Engineering applications of microworld physics" (No.~CZ.02.1.01/0.0/0.0/16\_019/0000766). The numerical calculations of this paper were performed using the computer Oakforest-PACS of the Joint Center for Advanced High Performance Computing through the program of the High Performance Computing Infrastructure in the fiscal year 2020 (hp200001) and the Multidisciplinary Cooperative Research Program 2020 of Center for Computational Sciences, University of Tsukuba (xg18i006).
\end{acknowledgments}

\bibliography{SM_QRPA9a}
\end{document}